\newif\ifBASIC
\BASICfalse
\newif\ifWP
\WPfalse

\WPtrue

\ifBASIC
  \documentclass[10pt]{article}
  \usepackage{amsmath,amsfonts,amssymb,amsthm,latexsym,graphicx,epigraph}
\fi

\ifWP
  \documentclass[12pt]{article}
  \usepackage{amsmath,amsfonts,amssymb,amsthm,latexsym,graphicx,epigraph,setspace}

\makeatletter

\newif\iftwodates
\twodatesfalse

\renewcommand\maketitle{\begin{titlepage}%
  \let\footnotesize\small
  \let\footnoterule\relax
  \let \footnote \thanks
  \null\vfil
  \vskip 30\p@
  \begin{center}%
    {\LARGE \bf \@title \par}%
    \vskip 3em%
    {\large
     \lineskip .75em%
     \begin{tabular}[t]{c}%
       \@author
     \end{tabular}\par}%
     \vskip 1.5em%
  \end{center}\par
  \vfill
  \begin{center}
    \raisebox{1.5cm}{\includegraphics[width=0.58\textwidth]%
      {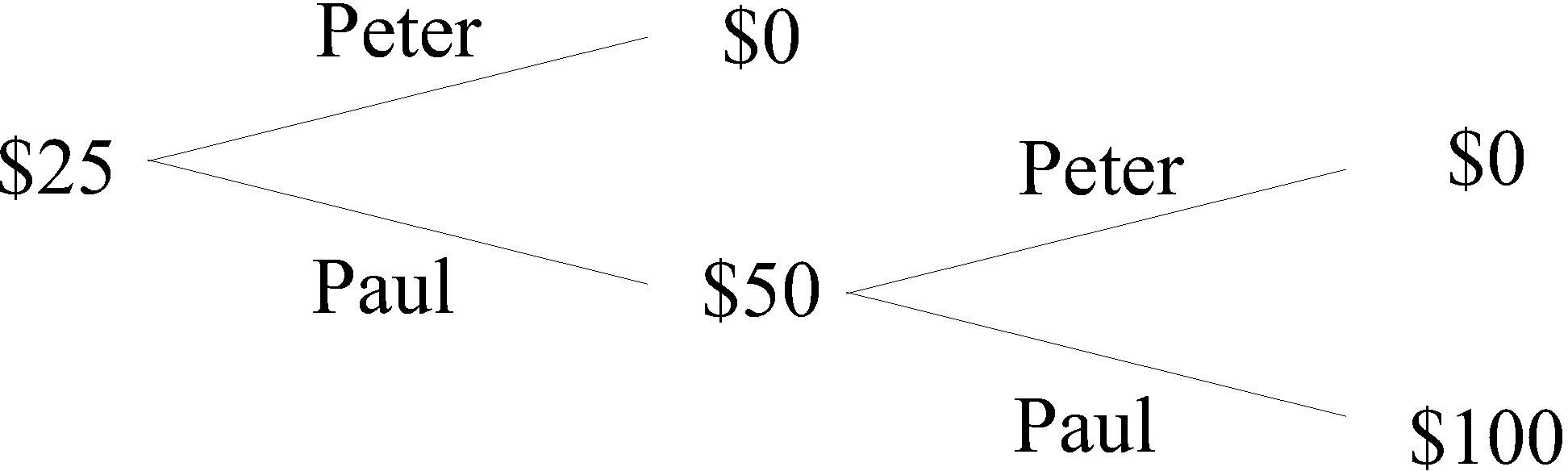}}%
    \hskip 3em%
    \includegraphics[width=0.29\textwidth]%
      {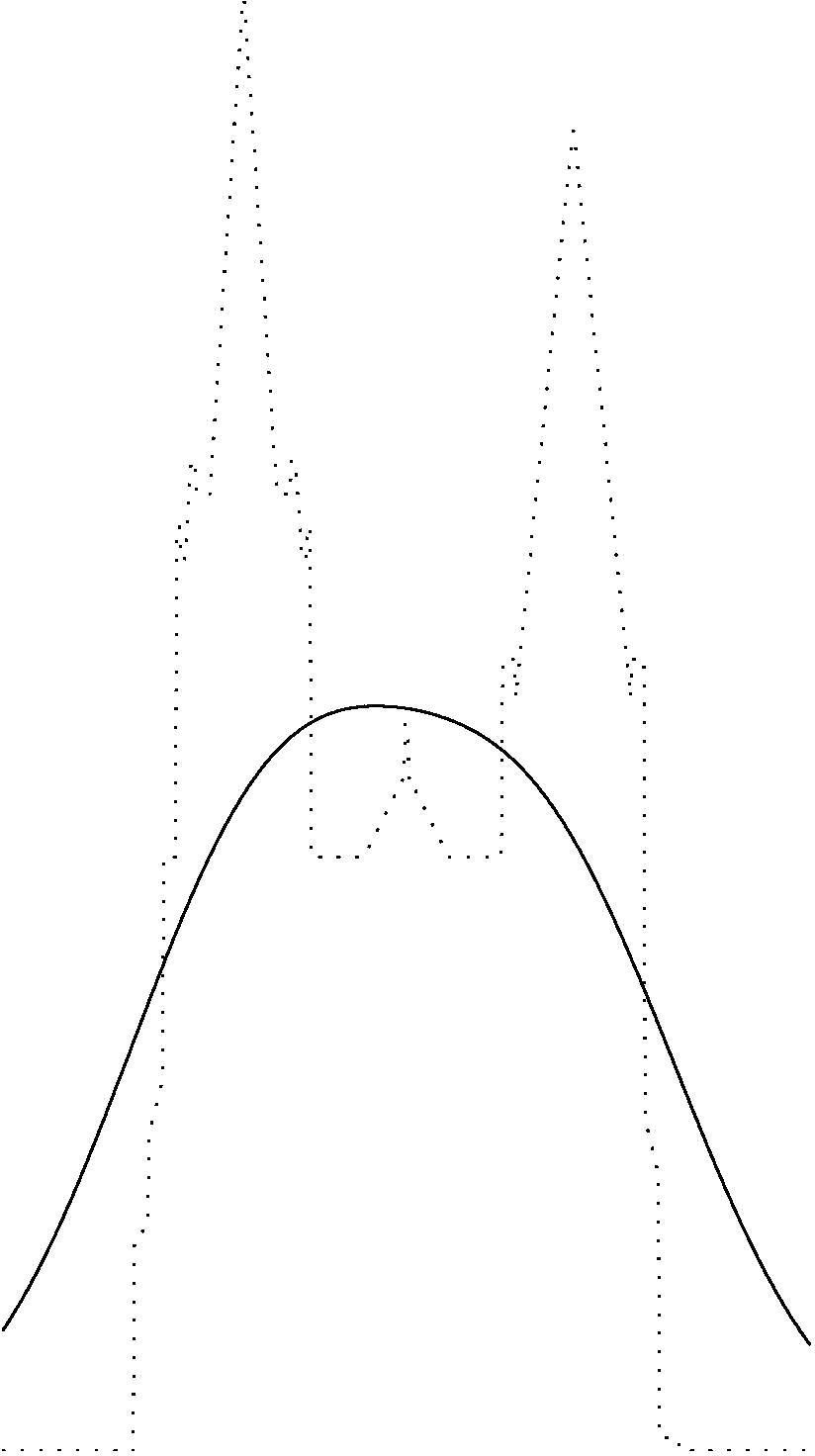}%
  \end{center}
  \@thanks
  \vfill
  \begin{center}
    {\large \bf The Game-Theoretic Probability and Finance Project}
  \end{center}
  \begin{center}
    {\large Working Paper \#\No}
  \end{center}
  \begin{center}
    {\iftwodates\large First posted \firstposted.
    Last revised \@date.\else\large\@date\fi}
  \end{center}
  \begin{center}
    Project web site:\\
    http://www.probabilityandfinance.com
  \end{center}
  \end{titlepage}%
  \setcounter{footnote}{0}%
  \global\let\thanks\relax
  \global\let\maketitle\relax
  \global\let\@thanks\@empty
  \global\let\@author\@empty
  \global\let\@date\@empty
  \global\let\@title\@empty
  \global\let\title\relax
  \global\let\author\relax
  \global\let\date\relax
  \global\let\and\relax
}

  {\par\vfill\tableofcontents\endtitlepage}

\renewenvironment{thebibliography}[1]
  {\section*{\refname}%
  \addcontentsline{toc}{section}{\refname}
  \@mkboth{\MakeUppercase\refname}{\MakeUppercase\refname}%
  \list{\@biblabel{\@arabic\c@enumiv}}%
    {\settowidth\labelwidth{\@biblabel{#1}}%
    \leftmargin\labelwidth
    \advance\leftmargin\labelsep
    \@openbib@code
    \usecounter{enumiv}%
    \let\p@enumiv\@empty
    \renewcommand\theenumiv{\@arabic\c@enumiv}}%
    \sloppy
    \clubpenalty4000
    \@clubpenalty \clubpenalty
    \widowpenalty4000%
    \sfcode`\.\@m}
    {\def\@noitemerr
    {\@latex@warning{Empty `thebibliography' environment}}%
  \endlist}

\makeatother

\fi

\allowdisplaybreaks[4]

\theoremstyle{plain}

\theoremstyle{definition}

\newcommand{\K}{\mathcal{K}}

\title{Kolmogorov's strong law of large numbers in game-theoretic probability: Reality's side}
\author{Vladimir Vovk}

\ifWP
  \author{Vladimir Vovk}
  \newcommand{\No}{40}
\fi

\begin{document}
\maketitle

\section{Statement}

This note describes a simple explicit strategy for Reality
whose existence is asserted in Theorem 4.1 (part 2) of \cite{shafer/vovk:2001} (p.~80).
We will be using the notation of \cite{shafer/vovk:2001}
complemented by $S_n:=x_1+\cdots+x_n$;
without loss of generality we can assume that $m_n=0$ for all $n$.
Namely, we construct an explicit strategy for Reality that guarantees
\begin{equation}\label{eq:goal}
  \K_n \text{ is bounded and }
  \left(
    \sum_{n=1}^{\infty}
    \frac{v_n}{n^2}
    =
    \infty
    \Longrightarrow
    \left(
      \frac{S_n}{n} \not\to 0
    \right)
  \right)
\end{equation}
provided Skeptic satisfies his collateral duty (keeping $\K_n$ non-negative).

For much more advanced results, see Theorems~4.12
and~5.10 of \cite{miyabe/takemura:2012}.
The main advantage of this note is its brevity.

\section{Reality's strategy and proof}

With Skeptic's move $(M_n,V_n)$ we associate the function $f_n(x):=M_nx+V_n(x^2-v_n)$;
the increase in his capital will be $f_n(x_n)$.
We will assume that $M_n=0$:
it will be clear that Reality can exploit $M_n\ne0$ by choosing the sign of $x_n$.
Our argument will also be applicable to the modified protocol of unbounded forecasting
in which Skeptic can choose any $V_n\in\mathbb{R}$:
Reality can easily win when $V_n<0$ by choosing $\lvert x_n\rvert$ large enough.

This is Reality's strategy:
\begin{enumerate}
\item\label{it:loop}
  Keep setting $x_n:=0$
  until Skeptic chooses a move for which $\K_{n-1}+f_n(n)\le1$.
\item\label{it:end}
  When Skeptic chooses such a move, set $x_n:=n$ or $x_n:=-n$.
  Go to \ref{it:loop}.
\end{enumerate}
Notice that Skeptic's capital is guaranteed to be bounded by 1,
and so Reality satisfies her collateral duty.
Consider two cases:
\begin{itemize}
\item
  Suppose that item \ref{it:end} is reached infinitely often.
  Since $S_n/n \not\to 0$ as $n\to\infty$,
  (\ref{eq:goal}) will be satisfied.
\item
  Now suppose Skeptic reaches item \ref{it:end} only finitely many times.
  From some $n$ on, we will have $f_n(n)>1-\K_{n-1}$,
  i.e., $V_n(n^2-v_n)>1-\K_{n-1}$,
  which implies $V_n>n^{-2}(1-\K_{n-1})$.
  Therefore, from some $n$ on Skeptic will lose at least
  $\lvert f_n(0)\rvert=V_nv_n\ge v_nn^{-2}(1-\K_{n-1})$.
  Suppose, without loss of generality, $\sum_nv_nn^{-2}=\infty$
  (otherwise (\ref{eq:goal}) is satisfied).
  Since the sequence $1-\K_{n-1}$ is increasing from some $n$ on
  and there are arbitrarily large $n$ for which $v_n>0$ and hence $1-\K_n>1-\K_{n-1}\ge0$,
  Skeptic will eventually become bankrupt.
\end{itemize}

\subsection*{Acknowledgement}

This note was prompted by Akimichi Takemura's question in 2006;
I am also grateful to him for several useful discussions.

\end{document}